\documentclass[
reprint,
superscriptaddress,
a4paper,
 nofootinbib,
 nobibnotes,
 amsmath,amssymb,
 aps,
 prl,
floatfix,
]{revtex4-2}

\usepackage{graphicx}
\usepackage[caption=false]{subfig}
\usepackage{dcolumn}
\usepackage{bm}
\usepackage{siunitx}
\usepackage{multirow}
\usepackage{amsmath}
\usepackage{natbib}
\usepackage{hyperref}

\let\svthefootnote\thefootnote
\newcommand\freefootnote[1]{%
  \let\thefootnote\relax%
  \footnotetext{#1}%
  \let\thefootnote\svthefootnote%
}

\usepackage{xcolor}
\usepackage{soul}

\newcommand{\NiS}{\texorpdfstring{NiS$_2$}{NiS2}}

\usepackage[left=2cm,right=2cm,top=2cm,bottom=2cm]{geometry}

\makeatletter

\def\@fnsymbol#1{\ensuremath{\ifcase#1\or \dagger\or \ddagger\or \mathsection\or
   \|\or *\or \mathparagraph\or **\or \dagger\dagger
   \or \ddagger\ddagger \else\@ctrerr\fi}}

\renewcommand{\fnum@figure}{\textbf{Figure~\thefigure}}

\renewcommand\paragraph{%
  \@startsection
    {paragraph}%
    {4}%
    {0mm}%
    {0.5\baselineskip}%
    {-1em}%
    {\normalfont\normalsize\bf}%
}%
\makeatother

\renewcommand{\thefootnote}{\alph{footnote}}

\begin{document}

\title{Truncated mass divergence in a Mott metal}

\freefootnote{K.S., H.C., and J.B. contributed equally to this work}

\author{Konstantin Semeniuk}
\altaffiliation{Now at Max Planck Institute for Chemical Physics of Solids, Dresden 01187, Germany \footnote{test}}
\affiliation{Cavendish Laboratory, University of Cambridge, Cambridge CB3 0HE, United Kingdom}

\author{Hui Chang}
\affiliation{Cavendish Laboratory, University of Cambridge, Cambridge CB3 0HE, United Kingdom}

\author{Jordan Baglo}
\altaffiliation{Now at Department of Physics, Universit\'e de Sherbrooke, Sherbrooke J1K 2R1, Canada}
\affiliation{Cavendish Laboratory, University of Cambridge, Cambridge CB3 0HE, United Kingdom}

\author{Sven Friedemann}
\affiliation{H H Wills Laboratory, University of Bristol, Bristol BS8 1TL, United Kingdom}

\author{Audrey Grockowiak}
\altaffiliation{Now at Brazilian Synchrotron Light Laboratory, Brazilian Center for Research in Energy and Materials, 13083-970 Campinas, São Paulo, Brazil}
\affiliation{National High Magnetic Field Laboratory, Tallahassee, FL 83810, USA}

\author{William A. Coniglio}
\affiliation{National High Magnetic Field Laboratory, Tallahassee, FL 83810, USA}

\author{Monika Gamza}
\affiliation{Jeremiah Horrocks Institute for Mathematics, Physics and Astronomy, University of Central Lancashire, Preston PR1 2HE, UK}

\author{Pascal Reiss}
\altaffiliation{Now at Max Planck Institute for Solid State Research, Stuttgart 70569, Germany}
\affiliation{Cavendish Laboratory, University of Cambridge, Cambridge CB3 0HE, United Kingdom}
\affiliation{Clarendon Laboratory, University of Oxford, Oxford OX1 3PU, United Kingdom}

\author{Patricia Alireza}
\affiliation{Cavendish Laboratory, University of Cambridge, Cambridge CB3 0HE, United Kingdom}

\author{Inge Leermakers}
\affiliation{High Field Magnet Laboratory (HFML-EMFL), Radboud University, Toernooiveld 7, 6525 ED Nijmegen, The Netherlands}

\author{Alix McCollam}
\affiliation{High Field Magnet Laboratory (HFML-EMFL), Radboud University, Toernooiveld 7, 6525 ED Nijmegen, The Netherlands}

\author{Stan Tozer}
\affiliation{National High Magnetic Field Laboratory, Tallahassee, FL 83810, USA}

\author{F. Malte Grosche}
\email[Corresponding Author: ]{fmg12@cam.ac.uk}
\affiliation{Cavendish Laboratory, University of Cambridge, Cambridge CB3 0HE, United Kingdom}

\date{\today}

\maketitle

\noindent
{\bf Metal-insulator transitions in clean, crystalline solids can be driven by two distinct mechanisms. In a conventional insulator, the charge carrier concentration vanishes, when an energy gap separates filled and unfilled electronic states. In the established picture of a Mott insulator \cite{brinkman70}, by contrast, electronic interactions cause coherent charge carriers to slow down and eventually stop in electronic grid-lock without materially affecting the carrier concentration itself. This description has so far escaped experimental verification by quantum oscillation measurements, which directly probe the velocity distribution of the coherent charge carriers. By extending this technique to high pressure we were able to examine the evolution of carrier concentration and velocity in the strongly correlated metallic state of the clean, crystalline material NiS$_2$, while tuning the system towards the Mott insulating phase. Our results confirm that pronounced electronic slowing down indeed governs the approach to the insulating state. However, the critical point itself, at which the carrier velocity would reach zero and the effective carrier mass diverge, is concealed by the insulating sector of the phase diagram. In the resulting, more nuanced view of Mott localisation, the inaccessibility of the low temperature Mott critical point resembles that of the threshold of magnetic order in clean metallic systems, where criticality is almost universally interrupted by first order transitions \cite{brando16}, tricritical behaviour \cite{friedemann18} or novel emergent phases such as unconventional superconductivity \cite{mathur98}.}

Mott localisation is one of the most fundamental consequences of electronic interactions in solids \cite{imada98}. Its theoretical understanding feeds into numerous related research areas, ranging from cuprate superconductivity to Kondo lattice materials and correlated topological insulators \cite{lee06,dzero16}. The Mott insulating state  \cite{mott49} is stabilised near half-filling, when the on-site repulsion energy $U$ exceeds a threshold value $U_c$. In the simplest case laid out in the Hubbard model \cite{hubbard63}, this threshold is determined by the kinetic energy contribution to the total energy. Various factors further affect $U_c$, for instance charge transfer into additional bands near the Fermi energy or more than one half-filled state per lattice site and the resulting Hund's rule coupling. Although quantum materials of current interest often require more elaborate models that reflect, for instance, the interplay of slow and fast carriers in Kondo lattice systems, key aspects of these materials connect back to the Hubbard model, such as the notion of orbitally selective Mott transitions in multiband systems \cite{vojta10,si01,senthil04,demedici14}.

\begin{figure}
\centerline{\includegraphics[width=\columnwidth]{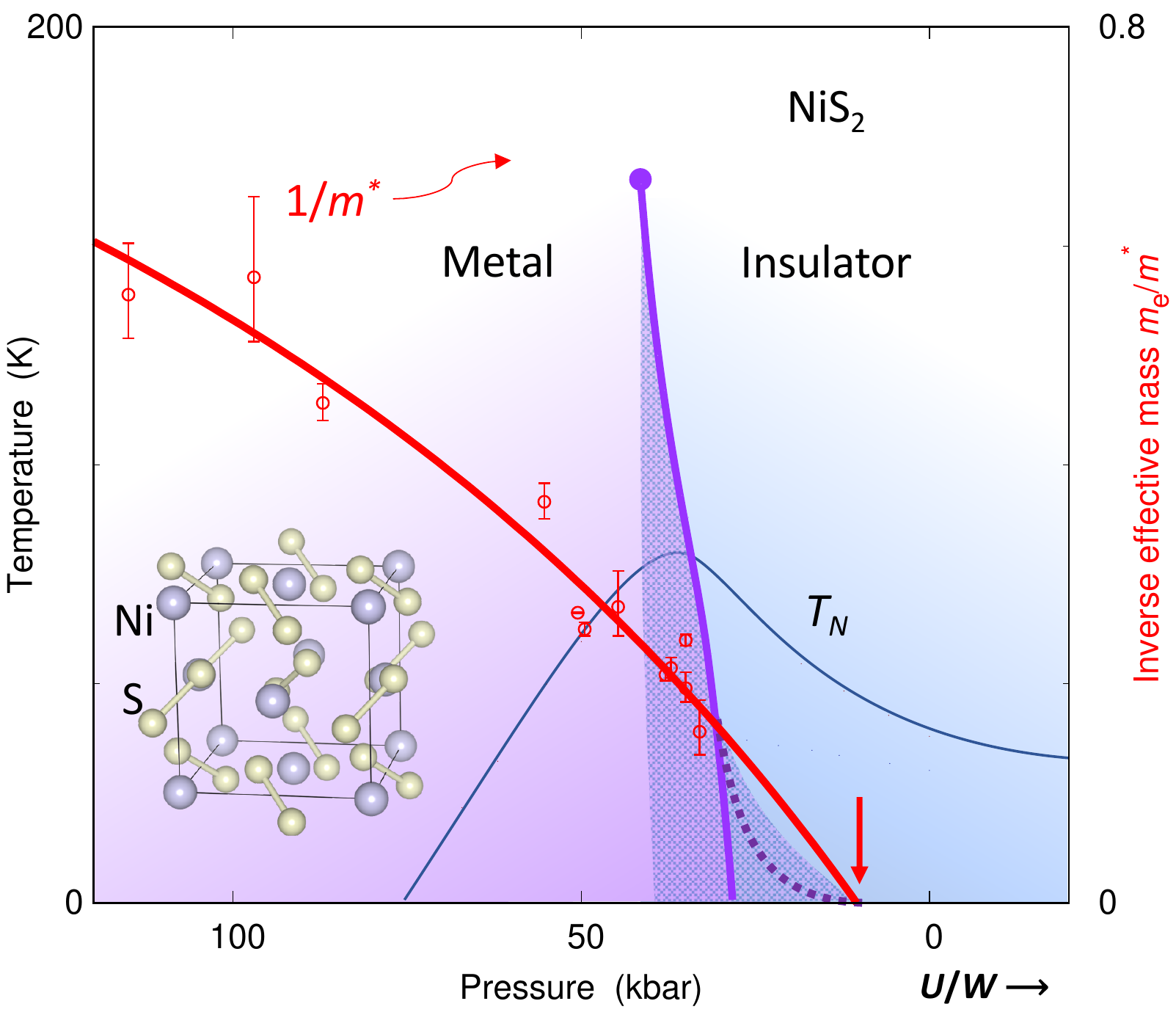}}
\caption{ {\bf Mott metal insulator transition in NiS$_2$,} which is metallic at high pressure (large effective bandwidth $W$ and therefore small ratio $U/W$) and insulating at low pressure. Magnetic order sets in below a transition temperature $T_N$ (blue line, following experimental data \cite{mori78, takeshita07}). The Mott transition line (purple, from resistivity data \cite{friedemann16}) ends in a critical point at high temperature. Its first-order nature implies the possibility of metastable states (dotted region) surrounding the thermodynamic transition line. At low temperature, the transition line in DMFT calculations curves away towards higher $U/W$ (dashed purple line), ending in a zero-temperature critical point \cite{georges96,georges04,kotliar04} (thick red arrow). The inverse carrier mass extracted in this study (red circles with errorbars) extrapolates to zero deep inside the region of the phase diagram where transport measurements show insulating behaviour.
}\label{fig:PhaseDiag} 
\end{figure}

In the canonical description formulated by Brinkman and Rice \cite{brinkman70}, Mott localisation is driven not by reduction of charge carrier concentration (as is the case in band insulators) but rather by a gradual slowing down of the charge carriers, while the volume enclosed by the Fermi surface remains constant in line with Luttinger's theorem \cite{luttinger60}. In this description, the reduction of the Fermi velocity of strongly correlated Landau quasiparticles is reflected in a reduction of the quasiparticle weight $z$ towards zero and a concomitant rise and, ultimately, divergence of the quasiparticle effective mass $m^*$. More sophisticated calculations within dynamic mean field theory (DMFT) have supported this scenario for the evolution of the correlated metallic state in the low temperature limit for a purely electronic Mott transition \cite{georges93,moeller95,georges96,georges04,kotliar04}. As illustrated in \autoref{fig:PhaseDiag}, these calculations indicate that the transition is first order at finite temperature, that it is accompanied by a range in which metallic and insulating states can coexist (dotted region in \autoref{fig:PhaseDiag}) and that the transition line bends sharply as the zero temperature limit is approached (dashed line in \autoref{fig:PhaseDiag}).

Given the central importance of Mott physics for understanding quantum materials it may be surprising that experimental tests of the Brinkman-Rice paradigm are comparatively scarce. Spectroscopic measurements such as photoemission spectroscopy (PES) \cite{Fujimori92,Inoue95} examine the suppression of the quasiparticle weight in the strongly correlated metal near Mott localisation, and they can track the spectrum close to and well away from the Fermi energy \cite{jang21}. However, the limited energy resolution of PES and its inability to distinguish between the coherent and incoherent parts of the spectrum hinder high precision measurements of the coherent low-energy excitations that constitute long-lived Landau quasiparticles, as is apparent for instance in \cite{xu14}. PES also tends to be limited to elevated temperatures, and it cannot be undertaken under pressure, which presents a clean way to tune in small steps across a Mott transition without doping-induced disorder. 

Conversely, the Landau quasiparticles at the core of the Brinkman-Rice picture are detected directly by observing quantum oscillatory phenomena in high magnetic fields. These probe the quasiparticle spectrum, making it possible to track the quasiparticle mass and Fermi surface as Mott localisation is approached with superior resolution. Quantum oscillation measurements have already contributed to reports of a divergent form of the quasiparticle mass at the \emph{magnetic} quantum critical points in CeRhIn$_5$ and BaFe$_2$(As$_{1-x}$P$_x$)$_2$ \cite{shishido05,hashimoto12}.  

Here, we study the clean Mott insulator \NiS\ and use high-pressure quantum oscillation measurements as a direct probe of the coherent quasiparticles and their Fermi velocity. 
Our data confirm key tenets of the Brinkman-Rice picture, namely that charge carrier concentration is conserved but carrier mass takes on a divergent form on approaching Mott localisation. In contrast to the Brinkman-Rice picture, however, we find that the carrier mass divergence extrapolates to a critical point buried well inside the insulating state. 

The cubic sulphide NiS$_2$ offers an excellent opportunity to investigate the correlated Mott metal, because high purity single crystals are available \cite{yao94,takeshita07} and a moderate pressure of about $\SI{30}{\kilo\bar}$ is sufficient to reach the metallic state \cite{wilson71,sekine97,takeshita07,friedemann16} (\autoref{fig:PhaseDiag}). A previous study demonstrated that quantum oscillations can be observed under these conditions \cite{friedemann16}. Avoiding doping and the associated disorder, this affords a direct view on the evolution of quasiparticle properties within the correlated metallic state. 

\begin{figure*}
\includegraphics[width=\textwidth]{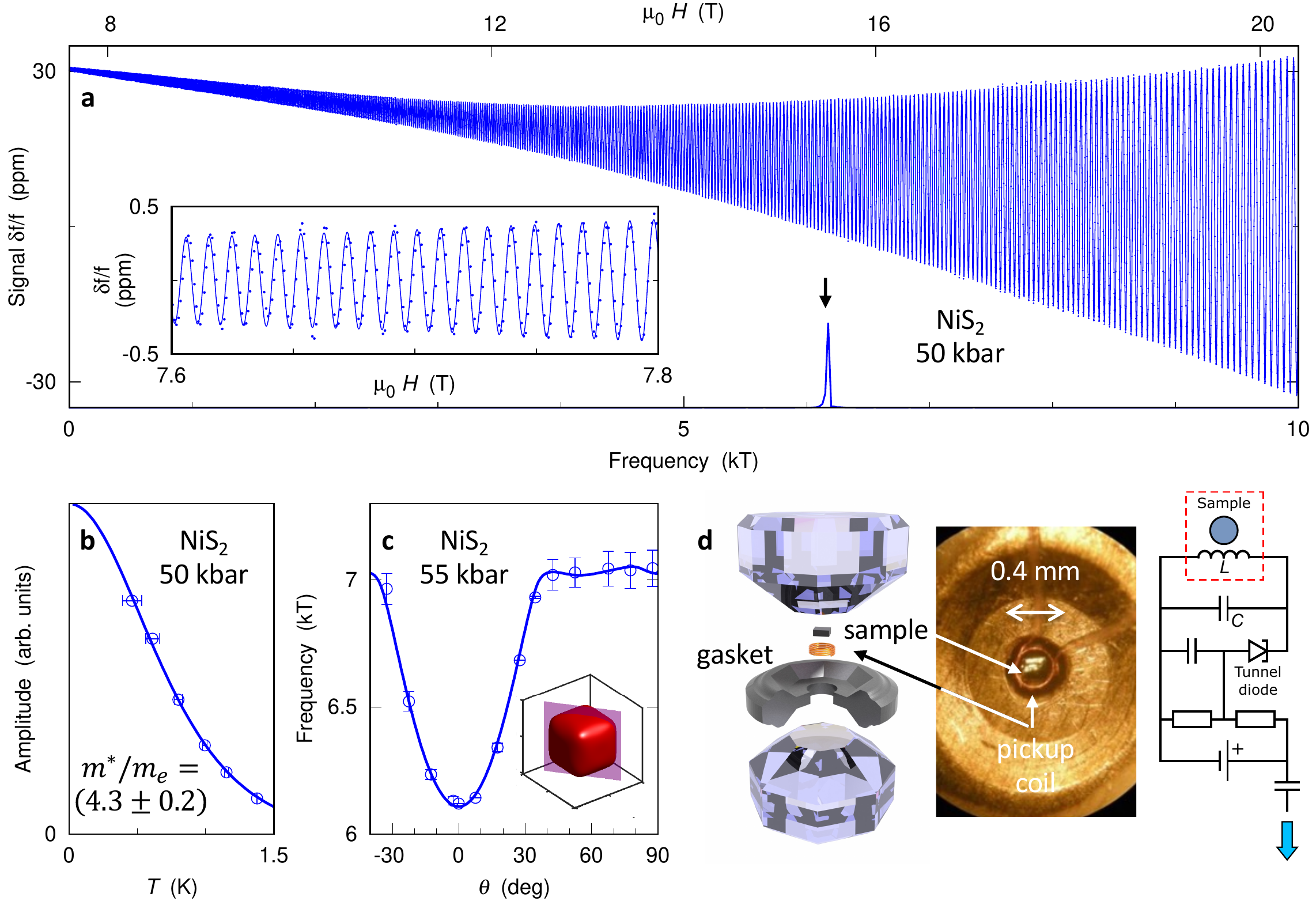}
\caption{ {\bf Quantum oscillations in NiS$_2$}. {\bf a,} Quantum oscillations are clearly resolved well above the metallisation pressure, down to fields as low as $<\SI{8}{\tesla}$ (inset). The power spectrum shows a single peak at $\SI{6.17}{kT}$ (lower axis). {\bf b,c,} The QO amplitude ({\bf b}) follows the Lifshitz-Kosevich form as a function of temperature (solid line), with effective mass $m^*\simeq 4.3~m_e$, and a rotation study ({\bf c}) at a nearby pressure produces an angle dependence of the QO frequency that closely matches expectations (solid line) for a  cube-shaped FS pocket (inset). {\bf d,} Key elements of the experimental setup.
}
\label{fig:QOMain}
\end{figure*}

The detection of quantum oscillations under pressure in NiS$_2$ \cite{friedemann16} builds on three innovations : (i) Te-flux growth \cite{yao94} reliably produces high quality crystals matching the best vapour-transport grown NiS$_2$ reported previously \cite{takeshita07}, with residual resistivities of $\simeq 1~\mu\Omega\text{cm}$ at high pressure \cite{friedemann16}; (ii) tank circuit radio frequency tunnel diode oscillator (TDO) techniques \cite{vandegrift75,goetze21} with a microcoil placed inside the $< \SI{400}{\micro\metre}$ diameter high pressure sample space enable ultra-sensitive, contact-free measurements of skin depth or magnetic susceptibility oscillations in applied field at high pressures (\hyperref[fig:QOMain]{Figure~2d}); (iii) miniature anvil cells and novel gasket preparation techniques allow access to the $\SI{100}{\kilo\bar}$ range and beyond in narrow-bore dilution refrigerator probes or on rotator stages while preserving excellent pressure homogeneity \cite{graf11}. Combined with superheterodyne signal detection \cite{suppmattemp}, such a setup can resolve quantum oscillations at the level of 0.01~ppm.

\begin{figure*}
\centerline{\includegraphics[width=\textwidth]{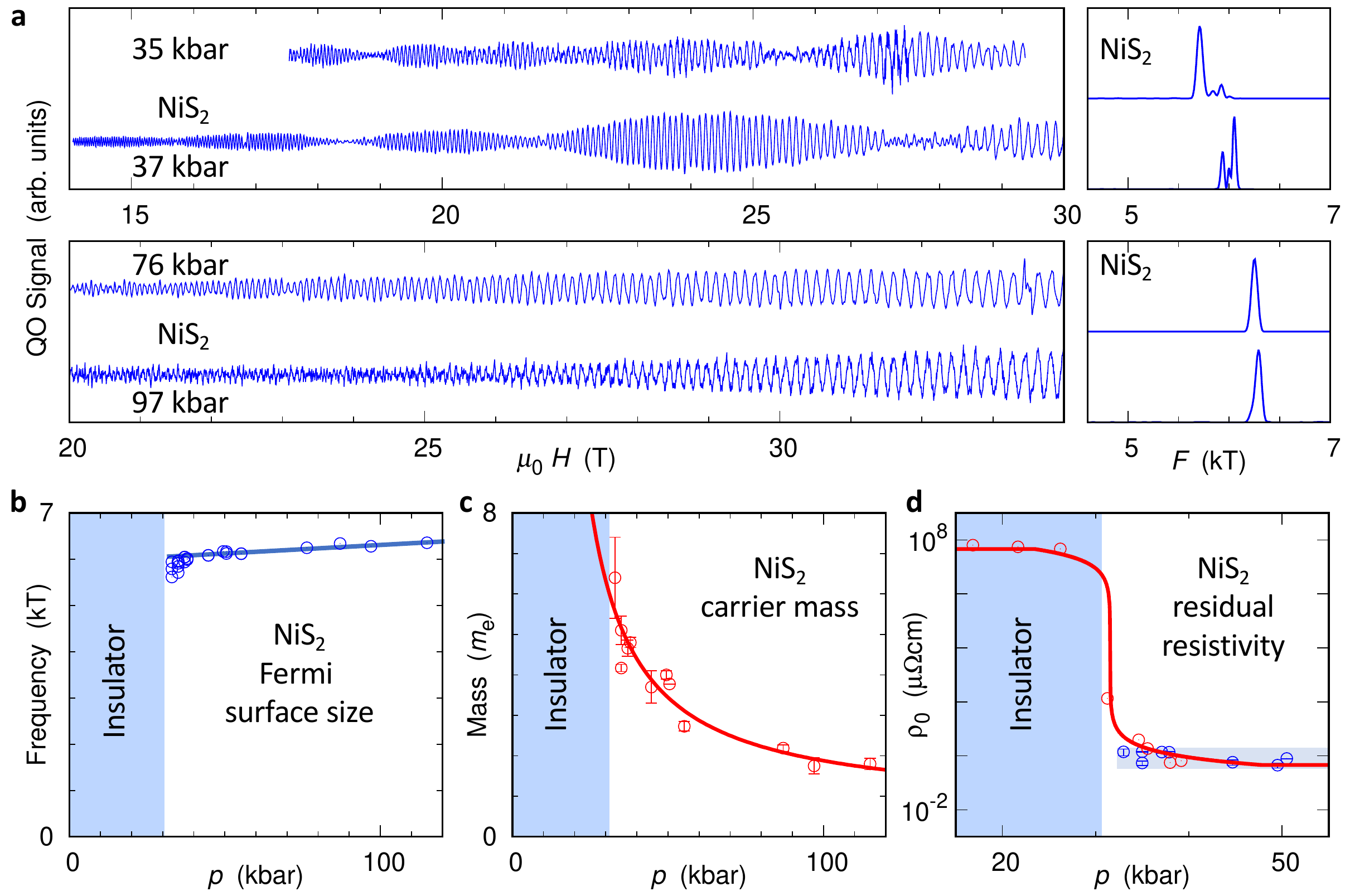}}
\caption{ {\bf Pressure evolution of the electronic strucutre of NiS$_2$.} {\bf a,} Quantum oscillations in NiS$_2$ at selected pressures (scaled and off-set for clarity), with corresponding power spectra to the right. {\bf b-d,} Pressure dependences of the quantum oscillation frequency ({\bf b}), the effective quasiparticle mass ({\bf c}) and the residual resistivity $\rho_0$ ({\bf d}). The blue line in panel {\bf b} is a linear least-squares fit to the frequency coming from the cubic Fermi surface (highest frequency at each pressure). The red line in panel {\bf c} is a least-squares fit consistent with the one shown in \autoref{fig:PhaseDiag} (see text). Whereas the effective mass is enhanced by about a factor of three over the investigated pressure range on approaching Mott localisation, the frequency and thereby the Fermi surface cross-section changes only slightly, at a rate which is broadly consistent with the change of the unit cell volume. Panel {\bf d} compares $\rho_0$ determined in high pressure transport measurements \cite{friedemann16} (red markers), to $\rho_0$ expected from the electronic mean free paths determined from the quantum oscillation analysis (blue markers). Whereas the quantum-oscillation-derived mean free path and the resulting $\rho_0$ are pressure independent (blue shaded region), direct transport measurements show a strong increase of $\rho_0$ on approaching the metal-insulator threshold from high pressures. This discrepancy suggests a significant and increasing volume fraction of insulating regions close to the metal-insulator threshold, which we model within 3D effective medium theory (red line). We ascribe the saturation of the transport-derived $\rho_0$ below \SI{30}{\kilo\bar} to surface conduction \cite{Clark16}.}
\label{fig:MassFreqPlot}
\end{figure*}

\paragraph{High-pressure quantum oscillations\\}
The Fermi surface signatures  detected in our quantum oscillation measurements in NiS$_2$ closely match density functional theory (DFT) calculations for the largest Fermi surface sheet  \cite{suppmattemp}. We observe strong quantum oscillations with a frequency $F \simeq \SI{6.17}{kT}$ as illustrated in (\autoref{fig:QOMain}) at a pressure of $\SI{50}{\kilo\bar}$, well above the metallisation pressure of $\SI{30}{\kilo\bar}$. This frequency is in close agreement with our earlier measurements at $\SI{38}{\kilo\bar}$ and with the accompanying DFT calculations \cite{friedemann16}. It corresponds to a cross-sectional area $A_k=\SI{0.589}{\angstrom}^{-2}$ for $B||c$, nearly half the cross-sectional area of the first Brillouin zone (BZ) $A_{BZ} = \SI{1.27}{\angstrom}^{-2}$. The angular dependence of the quantum oscillation frequency closely matches expectations from a cube-shaped Fermi surface pocket (\hyperref[fig:QOMain]{Figure~2c}). A hole pocket of almost identical size and shape is the dominant feature in {\em ab initio} DFT calculations within the paramagnetic metallic state \cite{friedemann16}, confirming the assignment of these quantum oscillations to the cube-shaped hole surface. A small splitting of the quantum oscillation frequency at lower pressures (top panel in \hyperref[fig:MassFreqPlot]{Figure~3a}) can be attributed to the effect of magnetic ordering \cite{suppmattemp}.  Whereas {\em ab initio} DFT calculations can accurately reflect the Fermi surface geometry, they capture electronic correlations  insufficiently to produce reliable estimates of the true carrier mass, which can be detected directly in quantum oscillation measurements.

The carrier mass is strongly renormalised. The effective quasiparticle mass $m^*= 4.3 m_e$ ($m_e$ - free electron mass) determined at $\SI{50}{\kilo\bar}$ from the temperature dependence of the oscillation amplitude (\hyperref[fig:QOMain]{Figure~2b}) exceeds the {\em ab inito} band mass $m_b = 0.8m_e$ obtained for this hole pocket by a factor of 5.4, indicating substantial mass renormalisation comparable to the highest values observed by quantum oscillation measurements in any transition metal compound \cite{shibauchi14,baglo21}. This highlights the strongly correlated nature of the metallic state in \NiS. 

The Fermi surface volume is preserved over the full pressure range, indicating that the charge carrier concentration remains constant on approaching the metal-insulator transition. The pressure dependence of the quantum oscillation frequency and carrier mass, based on 14 high pressure runs reaching up to $\sim \SI{115}{\kilo\bar}$, are summarised in \autoref{fig:MassFreqPlot}. The pressure dependence of the quantum oscillation frequency can be attributed to the compressibility of \NiS\ as indicated by the solid line in \hyperref[fig:MassFreqPlot]{Figure~3b}. We use the change of unit cell volume determined from X-ray diffraction to estimate the effect of pressure on the Fermi-surface volume \cite{suppmattemp}.  Transport measurements in the Ni(S/Se)$_2$ composition series indicate increases in resistivity and Hall coefficient on approaching the insulating threshold by reducing the Se content, which could be interpreted in terms of a decreasing carrier concentration  \cite{miyasaka00}. Our data show that this scenario does not apply in pressure-metallised NiS$_2$: the nearly constant quantum oscillation frequencies observed right up to the insulating state demonstrate that no significant change in the carrier concentration is taking place. We note that quantum oscillations directly probe the Fermi surface volume in reciprocal space, whereas contrary inferences from Hall effect measurements \cite{miyasaka00} may suffer from the effects of inhomogeneities and magnetic contributions.

The carrier mass is strongly boosted on approaching the Mott metal-insulator transition, suggesting critical slowing down of the coherent quasiparticles. 
We observe a monotonic 3-fold increase of the quantum oscillation mass between $\SI{120}{\kilo\bar}$ and $\simeq\SI{30}{\kilo\bar}$ (\hyperref[fig:MassFreqPlot]{Figure~3c}). This rules out magnetic quantum criticality associated with the threshold of antiferromagnetism at  $\approx\SI{80}{\kilo\bar}$ \cite{niklowitz08} as the primary driver of the mass enhancement. The effective mass of the coherent quasiparticles in \NiS\ is enhanced up to seven-fold above the bare (DFT) band mass, which in our calculations shows negligible pressure dependence over the range investigated experimentally.

While the carrier mass follows a divergent form consistent with the Brinkman-Rice scenario over a wide range of pressure, the critical point for the mass divergence is buried within the insulating part of the phase diagram. The quasiparticle mass is expected to diverge as $1/(p-p_c)$ within the Brinkman-Rice picture close to a critical pressure $p_c$. More generally, the inverse mass enhancement given in \cite{brinkman70} follows $m_e/m^* = 1-\left(U/U_0\right)^2$, where $U_0$ is proportional to the band width and can be expanded to first order in pressure $p$: $U_0 = A (p+a)$, with constant coefficients $A$ and $a$. At the critical pressure $p_c$, $U_0 = A(p_c + a)= U$, leading to the simpler expression for the inverse mass renormalisation  $m_e/m^* = 1-\left((p_c+a)/(p+a)\right)^2$. This expression closely fits the pressure dependence of the quantum oscillation mass  (\hyperref[fig:MassFreqPlot]{Figure~3c}), with a critical pressure $p_c \simeq \SI{10\pm 1}{\kilo\bar}$ well below the metallisation pressure of about $\SI{30}{\kilo\bar}$ determined from transport experiments  \cite{friedemann16}. The inverse mass enhancement, likewise, is shown in \autoref{fig:PhaseDiag} to extrapolate to zero inside the insulating regime.

The resulting phase diagram (\autoref{fig:PhaseDiag}) is consistent with DMFT studies for a purely electronic Mott transition \cite{georges93,georges96,georges04,moeller95,kotliar04}. As mentioned in the introduction, the first-order thermodynamic transition line in these calculations bends sharply towards higher coupling at low $T$, and it is surrounded by a region in which which both metallic and insulating states coexist, one being thermodynamically stable, the other metastable. Below the pressure of the high $T$ critical point, $p_h \simeq \SI{44}{\kilo\bar}$, the low temperature state is accessed by cooling through the transition line which overhangs the metallic region of the phase diagram. If the thermodynamic transition line is crossed at sufficiently high temperature, enough of the material converts into the metallic state to enable the observation of quantum oscillations. The reduction in QO signal amplitude and the rise in $\rho_0$ on approaching the metal-insulator transition in NiS$_2$ \cite{friedemann16} (\hyperref[fig:MassFreqPlot]{Figure~3d}) as well as Ni(S/Se)$_2$  \cite{miyasaka00} suggest that the metallic volume fraction is already strongly reduced well before the metal insulator transition itself is reached. In this purely electronically driven scenario, the metallic volume fraction diminishes as the second-order low temperature  end point of the Mott transition is approached, truncating the observed mass divergence. Coupling to the lattice, which causes a discontinuous volume change, can further exacerbate the first-order nature of the MIT and extend it to zero temperature \cite{majumdar94}, folding away the section of the pressure-temperature phase diagram surrounding the second-order low temperature end point of the transition line, which thereby becomes inaccessible to experiment.  

This interpretation is consistent with high pressure x-ray data, which reported evidence for phase coexistence near metallisation in NiS$_2$ \cite{feng11}, with ARPES data in Ni(S/Se)$_2$, which indicates that the Fermi velocity $v_F$ extrapolates to zero within the insulating part of the phase diagram \cite{xu14,jang21}, and with high pressure heat capacity measurements in V$_2$O$_3$, in which the divergence of the Sommerfeld coefficient does not occur at the metal insulator transition but rather can be extrapolated well into the insulating phase \cite{Carter93}. 

In summary, our high pressure quantum oscillation measurements allow the first comprehensive survey of the electronic structure and its evolution on the metallic side of a pressure-induced Mott insulator transition in ultra-clean specimens with a simple crystal structure. A large Fermi surface pocket is detected in pressure-metallised NiS$_2$, which corresponds closely to expectations from band structure calculations. On approaching Mott localisation from the metallic side, the Fermi surface volume remains comparatively unaffected right up until metallic behaviour is lost. The carrier mass, on the other hand, rises in a clean, divergent form with unprecedented dynamic range for correlation-induced mass renormalisation. 

Our results complement the central tenets of the Brinkman-Rice picture with the realisation that in this cubic system the expected mass divergence on approaching Mott localisation is truncated by the consequences of a first order phase transition, reminiscent of the fate of {\em magnetic} quantum critical points in metallic magnets \cite{brando16}. Resolving the precise origin of this phenomenon may be a challenge for future experiments which could, for instance, approach the critical pressure from the metallic side at low temperature in a variable-pressure device. Our results furthermore demonstrate the power of anvil cell based quantum oscillation measurements extending into the $>\SI{100}{\kilo\bar}$ regime for addressing challenging questions in fundamental condensed matter research. This provides a powerful complement to ARPES studies, which are limited to ambient conditions, and opens up a wide range of long-standing problems for closer investigation. The metallic state on the threshold of a Mott insulator transition is a central research theme in modern condensed matter physics. Its relevance is born out by the intense effort devoted to the normal state of the high temperature superconducting cuprates. Additional complexity arises from coupling to the lattice \cite{majumdar94,georgescu21}, orbital degeneracy and Hund's rule coupling \cite{nevidomskyy09,georges13,jang21}, phase separation and percolation \cite{pustogow21}, colossal susceptibility to applied electric field or strain \cite{kim10,nakamura13,han18a} and novel electronic surface or bulk states \cite{hartstein18}. The methodology developed for this study may help investigate these and other challenging phenomena that arise in the correlated metallic state on the threshold of Mott localisation -- or more generally near pressure-induced quantum phase transitions.

\section{Methods}

\paragraph{Synthesis of NiS$_2$ single crystals.} Single-crystal samples were grown with the tellurium-flux method as described earlier \cite{yao94}. Ultra-low sulphur deficiency was determined from X-ray diffraction results as detailed in \cite{friedemann16}.

\paragraph{Tunnel-diode-oscillator measurements at high pressures.} Quantum oscillations were measured via a contactless skin-depth probing technique based on the tunnel diode oscillator (TDO)---an inductor-capacitor oscillator sustained by a tunnel diode \cite{vandegrift75}. The inductors in our experiments were 3-10 turn cylindrical coils with inner diameters of $\SIrange{80}{200}{\micro\metre}$ wound with $\SIrange{12}{15}{\micro\metre}$ insulated copper wires. The coils were mounted inside the $<\SI{400}{\micro\metre}$ diameter sample spaces of BeCu or composite \cite{graf11} gaskets of diamond/moissanite anvil cells. Samples of NiS$_2$ were obtained by cleaving oriented single crystals into rectangular cuboids of $\SIrange{10}{100}{\micro\metre}$ thickness and  $\SIrange{50}{120}{\micro\metre}$ length and width, and were placed inside the coils. Our setup monitored the resonance frequency of the TDO. Changes in resistivity or magnetic susceptibility of NiS$_2$ in the metallic state were detected as proportional shift in the resonance frequency, due to a change in the flux expulsion caused by the skin effect, which depends on resistivity and magnetic susceptibility  \cite{suppmattemp}. Wirings of the cryostats were optimised to allow the oscillator to operate at frequencies of up to $\SI{500}{\mega\hertz}$ \cite{suppmattemp}. To ensure good hydrostaticity we used the 4:1 methanol-ethanol mix or 7474 Daphne oil (only at 55 kbar) as pressure-transmitting media. Pressure was determined via ruby fluorescence spectroscopy at low temperature. Quantum oscillations measurements were carried out at the NHMFL, Tallahassee, and at the HFML, Nijmegen, in top-loading $^3$He and dilution refrigeration cryostats with magnetic field strengths of up to $\SI{35}{\tesla}$, and at the superconducting high field facility in the Cavendish Laboratory, Cambridge, using a dilution refrigerator insert and fields of up to $\SI{18.4}{\tesla}$.

\paragraph{Analysis of quantum oscillations.} The cross-sectional area $A_k$ associated with a cyclotron orbit in strong magnetic fields is determined via the Onsager formula \cite{shoenberg84} $A_k = \frac{2 \pi e}{\hbar} F$, where $e$ and $\hbar$ are the elementary charge and Planck's constant, respectively. The quasiparticle mass and the mean free path were obtained via the full Lifshitz-Kosevich formula fitting \cite{suppmattemp}. The quasiparticle mass $m^{*}$ was additionally extracted by fitting the temperature ($T$) dependence of the quantum oscillation amplitude $\tilde y$ at a fixed magnetic field $B$ with the Lifshitz-Kosevich expression $\tilde y = \alpha T \left[\sinh\left(\SI{14.639}{\tesla\kelvin^{-1}} \frac{T}{B} \frac{m^*}{m_e}\right)\right]^{-1}$, where $\alpha$ is a temperature-independent factor and $m_e$ is the bare electron mass.

\section{Additional information}
\paragraph{Acknowledgements.} We thank, in particular, A. Chubukov, G. Lonzarich, and M. Sutherland for helpful discussions. The work was supported by the EPSRC of the UK (grants no. EP/K012894 and EP/P023290/1), and by Trinity College. Portions of this work were performed at the National High Magnetic Field Laboratory, which is supported by the National Science Foundation Cooperative Agreement No. DMR-1157490 and DMR-1644779 and the State of Florida, and at HFML-RU/NWO-I a part of the European Magnetic Field Laboratory (EMFL), which is supported by the EPSRC of the UK via its membership to the EMFL (grant no. EP/N01085X/1).
\paragraph{Author contributions.}
K.S., H.C., and J.B. contributed equally to this work. S.F. and F.M.G. conceived the experiment. S.F. and M.G. grew and characterised the NiS$_2$ single crystals. H.C. designed the TDO setup. H.C., A.G., and P.A. closed the high-pressure anvil cells. K.S., J.B., H.C., S.F., P.R., and F.M.G. conducted the quantum oscillation measurements. A.G., W.A.C., and S.T. provided technical support at NHFML Tallahassee. I.L. and A.M. provided technical support at HFML Nijmegen. K.S., J.B., H.C., S.F., P.R, and F.M.G. analysed the experimental data. P.R. performed the band structure and effective medium theory calculations. K.S., S.F., P.R., and F.M.G. co-wrote the manuscript.

\paragraph{Competing Interests.} The authors declare no competing interests.

\paragraph{Correspondence and requests for materials} should be addressed to F. Malte Grosche (fmg12@cam.ac.uk).

\bibliography{NiS2Refs_ArXiv.bib}

\end{document}